# Sieving hydrogen isotopes through two dimensional crystals

M. Lozada-Hidalgo[1], S. Hu[1], O. Marshall[1], A. Mishchenko[1], A. N. Grigorenko[1], R. A. W. Dryfe[2], B. Radha[1], I. V. Grigorieva[1], A. K. Geim[1]

[1]School of Physics & Astronomy and [2]School of Chemistry, University of Manchester, Manchester M13 9PL, UK

One-atom-thick crystals are impermeable to all atoms and molecules but hydrogen ions (thermal protons) penetrate relatively easily through them. We show that monolayers of graphene and boron nitride can be used to separate hydrogen isotopes. Employing electrical measurements and mass spectrometry, we found that deuterons permeate through these crystals much slower than protons, resulting in a large separation factor of $\approx$10 at room temperature. The isotope effect is attributed to a difference of $\approx$60 meV between zero-point energies of incident protons and deuterons, which translates into the equivalent difference in the activation barriers posed by two dimensional crystals. In addition to providing insight into the proton transport mechanism, the demonstrated approach offers a competitive and scalable way for hydrogen isotope enrichment.

Unlike conventional membranes used for sieving atomic and molecular species, monolayers of graphene and hexagonal boron nitride (hBN) exhibit subatomic selectivity (*1-5*). They are permeable to thermal protons (*5*) – and, of course, electrons (*6*) – but in the absence of structural defects are completely impermeable to larger, atomic species (*1-5,7-10*). Proton transport through these two-dimensional (2D) crystals was shown to be a thermally activated process, and the found energy barriers $E$ of $\approx$0.3 and 0.8 eV for monolayers of hBN and graphene, respectively, were attributed to different densities of their electron clouds that have to be pierced by incident protons (*5*). Investigating whether deuterons – nuclei of the heavier hydrogen isotope, deuterium (D) – can pass through atomically-thin crystals is interesting for both elucidating the proton transport mechanism and exploring its potential for applications. Indeed, substituting protium (H) with D can be expected to shed light on the transport process in analogy to chemical reactions, where such substitution is conventionally used to probe the role of proton transfer (*11-14*). If, in addition, the 2D membranes can distinguish between the two nuclei (hydrons), this would be of interest for applications: Hydrogen isotopes are important for various analytical and tracing technologies whereas heavy water is used in huge quantities by nuclear fission plants. Yet, the current H/D separation techniques such as, for example, water-hydrogen sulfide exchange and cryogenic distillation (*15,16*) show low separation factors (<2.5) and are among the most energy-intensive processes in the chemical industry. The present situation stimulates continuous search for alternative technologies (*15-21*).

In this report, we have investigated whether deuterons ($D^+$) permeate through 2D crystals differently from protons ($H^+$) studied previously (*5*). This was done using two complementary approaches: electrical conductivity measurements and gas flow detection by mass spectrometry (*22*). In the first approach, graphene and hBN monocrystals were mechanically exfoliated and suspended over micrometer-sized holes etched in silicon wafers (*fig. S1*). To measure 2D crystals' hydron conductivity $\sigma$, both sides of the resulting membranes were coated with a proton conducting polymer – Nafion (*23*) – and electrically contacted using Pd electrodes that converted electron into hydron flow (inset of Fig. 1A). The measurements were performed in either $H_2$-Ar/$H_2O$ or $D_2$-Ar/$D_2O$ atmosphere in 100% humidity at room temperature. The different atmospheres turned Nafion into a proton (H-Nafion) or deuteron (D-Nafion) (*24*) conductor with little presence of the other isotope (*fig. S2*). For reference, similar devices but without 2D membranes were fabricated. The latter exhibited similar conductance, whether H- or D-Nafion was used, and it was typically 100 times higher than that found for devices incorporating 2D crystals. This shows that the series contribution to our device resistances from Nafion and Pd contacts could be neglected (*5,22*).

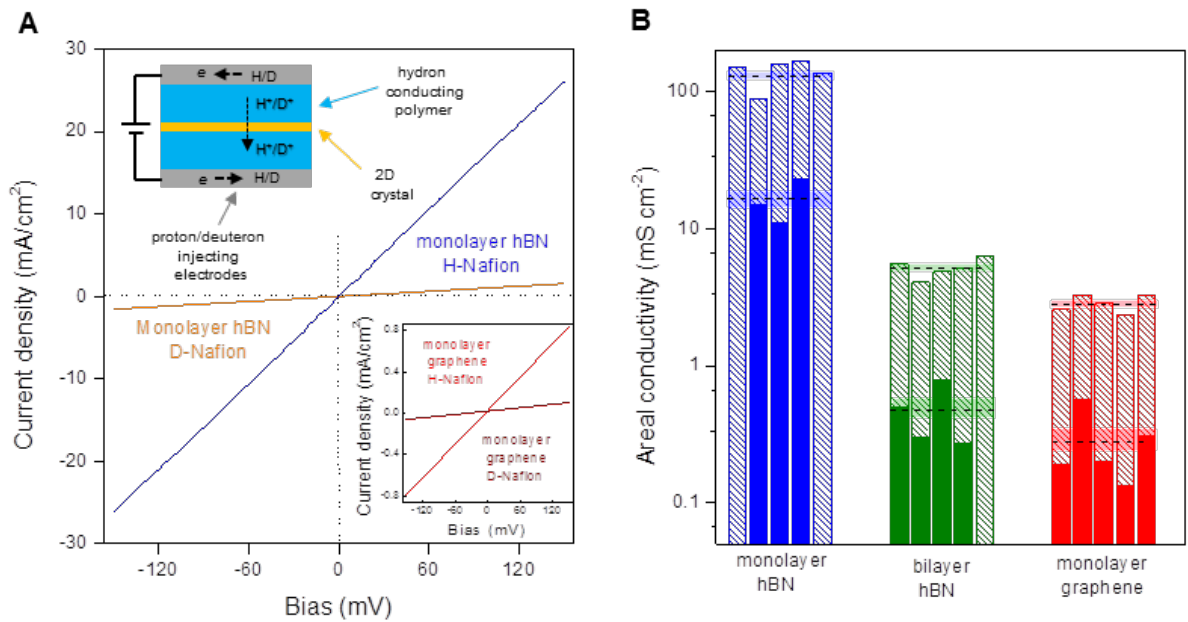

***Fig. 1. Proton vs deuteron conductivities of 2D crystals.*** *(**A**) Examples of I-V characteristics for hydron transport through monolayers of hBN and graphene. Top inset: Schematics of the experimental setup. Pd electrodes supply protons or deuterons into H- or D- Nafion; 2D crystals serve as barriers for hydrons. (**B**) Proton and deuteron conductivities (shaded and solid bars, respectively) for the most hydron conductive 2D crystals. Each bar (solid or shaded) corresponds to a different device (nearly thirty are shown). The dotted lines mark the average conductivities for the six sets of devices, and the shaded areas around them show the standard errors.*

For both H- and D- Nafion devices, the measured current $I$ varied linearly with applied bias $V$ (Fig. 1A) but different 2D crystals showed widely different areal conductivities, $\sigma = I/SV$ where $S$ is the membrane area (Fig. 1B). Monolayer hBN exhibited the highest proton conductivity $\sigma_H$, followed by bilayer hBN and monolayer graphene (Fig. 1B), in agreement with the previous work (*5*). Our main new finding is that $\sigma$ was markedly smaller (10 times) for D-Nafion devices compared to their H-Nafion counterparts, independently of the tested 2D crystal (Fig. 1B). Furthermore, we carried out similar measurements for Pt-activated membranes (hBN and graphene monolayers covered with a discontinuous layer of Pt to enhance hydron transport) (*22*) and, again, the conductivity for deuterons $\sigma_D$ was always ≈10 times lower than that for protons (*fig. S3*). This pronounced isotope effect is unexpected and its independence on 2D barrier height is particularly puzzling. These observations do not follow from either previous experiments (*5,10*) or existing theories (*7-9*) in which the calculated barriers arise due to the interaction of a positive point charge with the 2D crystals' electron clouds and the hydron mass is assumed irrelevant.

In our second series of experiments, graphene membranes were used to separate a liquid cell and a vacuum chamber (Fig. 2A). On the liquid side (input), graphene was coated with a thin Nafion layer that faced a reservoir containing a proton-deuteron electrolyte (HCl in $H_2O$ mixed with DCl in $D_2O$). The atomic fractions of $H^+$ and $D^+$ in this mixture could be changed as required. The other side of graphene, decorated with Pt nanoparticles, faced the vacuum chamber equipped with a mass spectrometer (*5,22*). A bias – typically, ≲2V to avoid damage to our devices (*fig. S6*) – was applied directly between graphene and the electrolyte (Fig. 2A; *fig. S1*). This setup effectively represents an electrochemical pump (*25,26*) in which the graphene membrane – impermeable to all gases and liquids – serves simultaneously as a semitransparent hydron barrier and a drain electrode for protons and deuterons. The aforementioned gas/liquid impermeability was confirmed for each experimental device by using a He leak detector. The key advantage of mass spectrometry with respect to our electrical measurements is that it can distinguish between different hydron species. This allowed us to determine directly the composition of output gas flows for different input electrolytes. Unfortunately, mass spectrometry is also much less sensitive than electrical measurements and, therefore, large hydron fluxes were necessary to probe the current-induced gas flows in the presence of a fluctuating background in the spectrometer (*22*). To compensate for the lower sensitivity, we used high *I* and graphene crystals as large as possible, fabricating membranes up to 50 μm in diameter. This allowed flows >>$10^{10}$ molecules/s for all three possible gases – $H_2$, $D_2$ and protium deuteride (HD) – which appeared on the vacuum side (Fig. 2B; *fig. S4*).

We found that the flow of each of the gases varied linearly with *I*, as expected, but depended strongly on the relative concentrations ([$H^+$]:[$D^+$]) of hydrons in the input electrolyte ([$H^+$]+[$D^+$] =100%). This is illustrated in Fig. 2B for the case of $D_2$ and further in *figs. S4-S6*. By measuring such flow-current dependences for different [$H^+$]:[$D^+$] inputs, we determined the percentage of $H_2$, $D_2$ and HD in output flows (Fig. 2C). These data are easily converted into the percentage of H and D atoms at the output of our electrochemical pump as a function of [$H^+$] or [$D^+$] at its input. Our main finding is that the output fraction of atomic hydrogen was disproportionally high with respect to the input fraction of protons, as shown in Fig. 2D. For example, for equal amounts of protons and deuterons at the input, H accounted for ≈95% of the atoms in the output flow, that is, graphene membranes efficiently sieved out deuterium. As a control experiment, we repeated the same measurements substituting graphene with porous carbon and found no preferential flow of either protons or deuterons, as expected (*fig. S5*). To quantify the observed sieving efficiencies we calculated the separation factor α. The data in Fig. 2D yield α ≈10, in good agreement with the value found from the ratio $\sigma_H/\sigma_D$ in our conductivity measurements (*22*).

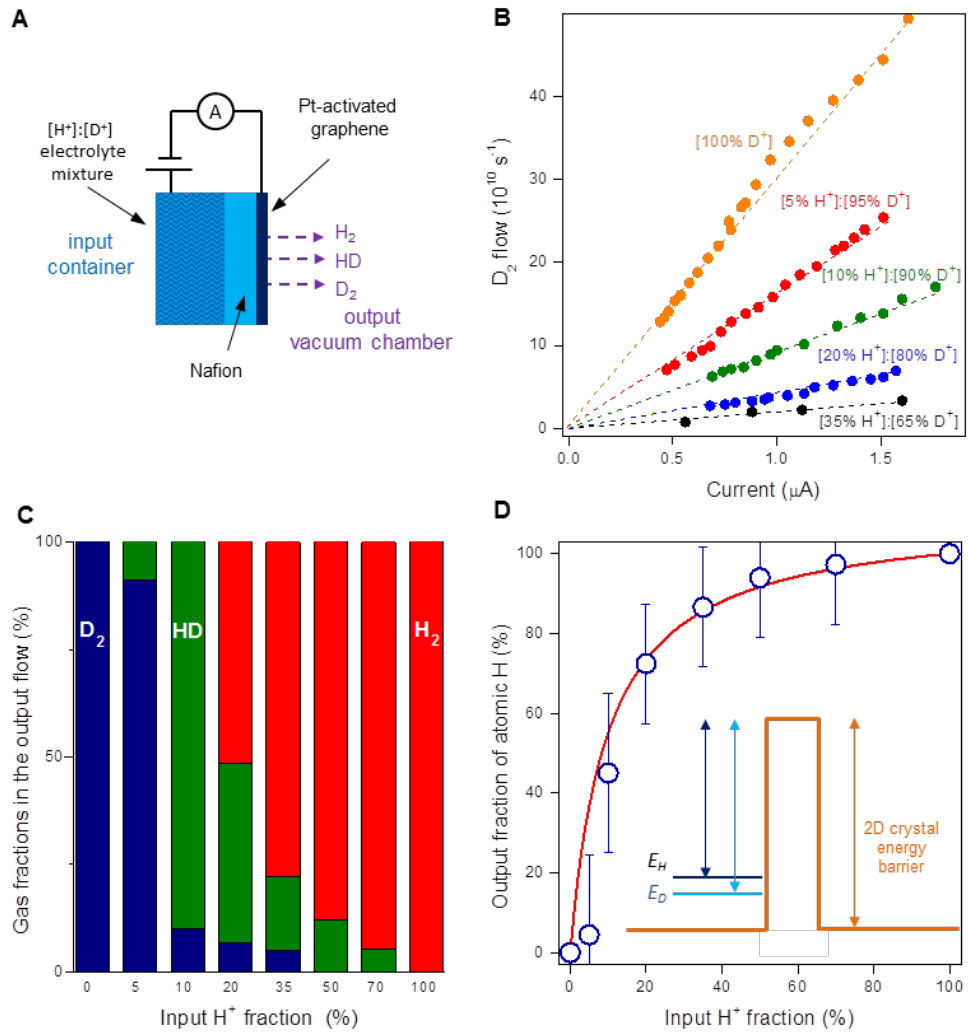

***Fig. 2. Isotope separation by electrochemical pumping of hydrons through graphene.*** *(**A**) Schematic of our mass spectrometry setup. (**B**) $D_2$ flow versus applied current for various proton-deuteron fractions in the input electrolyte. The dashed lines are linear fits. (**C**) Relative fractions of $H_2$, HD and $D_2$ in the output flow for eight different compositions of the input electrolyte. (**D**) Fraction of H atoms at the output for different [$H^+$] inputs. Inset: Schematic of the energy barrier presented by a 2D crystal for proton and deuteron transfer. The black and and blue horizontal lines indicate zero-point states of protons and deuterons, respectively, in Nafion and water. The solid red curve shows the separation dependence expected for the known difference $E_D - E_H$ = 60 meV, with no fitting parameters.*

To explain the observed isotope effect, we first recall that proton permeation through 2D crystals is a thermally activated process (*5,9*). Therefore, if we neglect – to a first approximation (*22*) – the pre-exponential factor in the Arrhenius equation, our results can equivalently be described in terms of the energy barriers $E_H$ and $E_D$ presented by 2D crystals to proton and deuteron transport, respectively. Accordingly, we can write $\sigma_H/\sigma_D = \exp(\Delta E/k_B T)$ where $\Delta E = E_D - E_H$. While $E_H$ and $E_D$ obviously determine the hydron permeability of 2D crystals (*5,7-9*), their selectivity depends only on *ΔE*. Statistical analysis of the data in Fig. 1B yields $\sigma_H/\sigma_D \approx 10 \pm 0.8$, which translates into *ΔE* ≈ 60±2 meV for all the tested 2D membranes. Furthermore, the same value of *ΔE* allows us to describe quantitatively the selectivity found by the mass spectrometry measurements. As shown in Supplementary Material, the protium output is given by $[H] = [H^+]/\{[H^+]+\exp(-\Delta E/k_B T)[D^+]\}$. This dependence is plotted in Fig. 2D using *ΔE* = 60 meV and shows excellent agreement with the experiment.

The above consideration leaves two questions: where does the difference in *ΔE* comes from and why is it the same for all the tested 2D membranes despite their hydron conductivities being different by many orders of magnitude? We point out that protons/deuterons in our experiments move not in vacuum but along hydrogen-bonded networks provided by sulfonate groups ($SO_3^-$) and water in Nafion (*23*). It is reasonable to expect that, before jumping through 2D crystals, hydrons remain transiently bonded to sulfonate/water groups and, accordingly, this presents the initial state in the transfer process (inset of Fig. 2D). The zero-point energies of these hydrogen-oxygen bonds are ≈0.2 eV for protons and ≈0.14 eV for deuterons (*11,22*). As illustrated in Fig. 2D, zero-point oscillations effectively reduce the activation barrier with respect to vacuum by 0.2 eV for protons whereas for deuterons the reduction is smaller by 60 meV. This explanation is consistent with all the experimental evidence, and the same-strength isotope effect is expected if the 2D crystals are combined with other proton conductors based on oxides (*13,25-28*). The effect could be even bigger for proton conducting media with stronger hydrogen bonds; for example, in fluorides (*28*).

The above explanation allows for several observations about proton transport through 2D crystals. First, it partially explains the disagreement between the experiment (*5*) and theory (*5,8-10*) in the absolute value of $E_H$ for graphene: zero-point oscillations reduce the activation barrier by ≈0.2 eV compared to theoretical values. We speculate that the remaining differences (<0.2 eV in the case of ref. 8) may be accounted for by considering other effects of the surrounding media (for example, two-body processes involving a distortion of the electron clouds by protons residing at the Nafion-graphene interface). Second, the experiments confirm that hydrogen chemisorption to 2D crystals is not the limiting step in the transfer process because, otherwise, the isotope effect would be different for hBN and graphene. Third, the described sieving mechanism implies α ≈30 for tritium-hydrogen separation. Fourth, it is quite remarkable that zero point oscillations, a purely quantum effect, can still dominate room-temperature transport properties of particles 4000 times heavier than electrons.

The observed large α compares favorably with sieving efficiencies of the existing methods for hydrogen isotope separation (*15-20*). Because graphene and boron nitride monolayers exhibit high proton conductivity, comparable to that of commercial Nafion films, (*5,22*) this makes them potentially interesting for such applications. In this respect, the increasing availability of graphene grown by chemical vapor deposition (CVD) (*29,30*) provides a realistic prospect of scaling up the described devices from micron sizes to those required for industrial uses. Indeed, while micromechanical cleavage allows 2D membranes of highest quality, the approach is not scalable. As a proof of concept, we repeated the mass spectrometry measurements using cm-size membranes made from CVD graphene and achieved the same α ≈10 (*fig. S7*). Importantly, this shows that macroscopic cracks and pinholes present in CVD graphene do not affect the efficiency because hydrons are electrochemically pumped only through the graphene areas that are electrically contacted (*22*). Furthermore, we estimate the energy costs associated with this isotope separation method as ≈0.3 kWh per kg of feed water (*22*), significantly lower than costs of the existing enrichment processes (*15,16*). All this comes on top of the fundamentally simple and robust sieving mechanism, potentially straightforward setups and only water at the input without the use of chemical compounds (*16*).

**Acknowledgments**: This work was supported by the European Research Council, the Royal Society and Lloyd's Register Foundation. M. L. H. and S. H. equally contributed to this work.

# Supplementary Materials

**Materials and Methods**

Electrical conductivity measurements

Devices for the study of hydron transport using electrical measurements were fabricated by suspending mechanically exfoliated 2D crystals over apertures that were etched in silicon-nitride membranes. The latter were made using standard Si wafers covered on both sides with 500 nm thick $SiN_x$ (*fig. S1A*). Graphite and hBN crystals were purchased from *NGS Naturgraphit* and *HQ Graphene*, respectively. The high sensitivity of electrical measurements allowed us to use relatively small holes (2 to 10 μm in diameter), and this allowed for studying not only graphene but also mono- and bi- layer hBN. This is because available hBN crystals cannot be exfoliated to monolayers with sizes similar to those achievable for graphene. Both sides of the suspended graphene and hBN membranes were coated with a Nafion layer, and either $PdH_x$ or $PdD_x$ electrodes (*31*) were mechanically attached to Nafion (see *figs. S1A-B*). We refer to section Methods of ref. *5* for more details on fabrication of such proton transport devices.

For electrical measurements, the assembled devices were placed in a chamber with a controlled atmosphere of either 10% $H_2$ in Ar at 100% $H_2O$ relative humidity or, alternatively, 10% $D_2$ in Ar at 100% $D_2O$ humidity. The reported I-V characteristics were measured with *Keithley*'s SourceMeter 2636A at voltages typically varying between ±200 mV and using sweep rates <0.1 V min$^{-1}$.

We first characterized our setup in terms of leakage currents and found parasitic parallel conductance of ~5 pS due to leakage along the silicon-nitride surface under humid conditions (*5*). In further control experiments, we measured the conductivity of D- and H- Nafion films using devices of the same design but without 2D crystal membranes. No difference could be found between Nafion enriched with the different isotopes, and D-Nafion bulk conductivity remained ~1 mS cm$^{-1}$, in agreement with the values reported previously for H-Nafion films prepared in the same manner (*5*).

Remnant protium content in D-Nafion

The vibrational modes of protons attached to water molecules have been studied using infrared spectroscopy (*32-34*). We employed this technique to estimate the amount of protons remaining in the D-Nafion films after their long exposure to $D_2O$. To this end, a Nafion solution was drop cast onto calcium fluoride windows to form films of ~1 μm in thickness. The windows were then integrated into an environmental chamber where the Nafion films were exposed to either 100% $H_2$ + $H_2O$ or 100% $D_2$ + $D_2O$ atmosphere. The measurements were carried out with a Bruker Vertex 80 FTIR spectrometer. Examples of the obtained spectra are shown in *fig. S2*.

H-Nafion exhibited a strong absorbance peak at ≈3,500 cm$^{-1}$, which corresponds to the stretching mode of the OH oscillator, $\nu_S$ (O-H). For D-Nafion, the corresponding mode is shifted by ≈1,000 cm$^{-1}$ to the frequency of the OD oscillator, $\nu_S$ (O-D). Importantly, only a very weak OH peak could be detected in D-Nafion (inset of *fig. S2*). By comparing its integrated intensity with the OH and OD peaks in the main panel, we estimate that, after changing the atmosphere from light to heavy water, the residual atomic fraction of H in D-Nafion was less than ≈1 %.

It is instructive to estimate the possible effect of the remnant protium found in our D-Nafion on the measured deuteron conductivity of the membranes. The measured areal conductivity for a D-Nafion device, $\sigma_D$, is given by $\sigma_D = 0.99\sigma_D^* + 0.01\sigma_H^*$ where $\sigma_D^*$ refers to the actual – rather than measured – deuteron areal conductivity and $\sigma_H^*$ is the areal conductivity associated with the residual protons in Nafion. Because deuterium concentration in H-Nafion is negligible (that is, $\sigma_H \equiv \sigma_H^*$) and the electrical measurements reported in the main text yielded $\sigma_H/\sigma_D \approx 10$, one can easily find that $\sigma_H^*/\sigma_D^* \approx 11$. Therefore, the correction due to H contamination of D-Nafion is relatively minor and within the statistical error of our transport measurements shown in Fig. 1B of the main text.

Pt decorated membranes

We also studied electrical conductivity of Pt-activated graphene and monolayer hBN. To prepare these membranes, Pt nanoparticles were deposited onto them by evaporating a discontinuous layer of Pt (nominally, 2 nm thick; see ref. 5). *Figure S3* shows examples of I-V characteristics for an hBN membrane prepared in this manner. As with non-decorated 2D crystals, we observed the same tenfold difference between the areal conductivities $\sigma$ of protons and deuterons.

The measurements in *fig. S3* also allow us to estimate room-$T$ proton conductivity for Pt-decorated monolayer hBN which exhibits the highest $\sigma$ among all the studied 2D crystal membranes. Previously (*5*), the proton conductivity of these hBN membranes was too high to be determined experimentally because H-Nafion devices, either with or without Pt-activated monolayer hBN, exhibited the same conductivity limited by Nafion. The deuteron transport measurements circumvent this problem. Because the bulk conductivities of H- and D- Nafion are the same, the observed deuteron conductivity for Pt-activated monolayer hBN in *fig. S3* yields its proton conductivity of $\approx 3$ S cm$^{-2}$. This is comparable with the areal conductivity of commercially available Nafion membranes of $\approx 200$ μm in thickness (*35*).

Mass transport measurements using exfoliated graphene

Graphene devices used in our mass spectrometry experiments are shown in *fig. S1C*. Similar to the devices for electrical measurements, they were fabricated by suspending monolayers of mechanically exfoliated graphene over apertures etched in silicon-nitride membranes (see above) but the apertures were much larger, 50 μm in diameter (*fig. S1D*). To achieve proton/deuteron flows sufficient for mass spectrometry detection, we catalytically activated graphene with the same discontinuous layer of Pt as described above. The Pt layer covered only the output (vacuum) side of the graphene membranes, which faced our mass spectrometer (Inficon UL200). The input side was coated with a Nafion film (5% solution; 1100 EW), and the assembly was then annealed in a humid atmosphere at 130°C to crosslink the polymer and improve its hydron conductance (*5*). As reference devices, we used the same assembly (*fig. S1C*) but graphene was substituted with a carbon cloth containing Pt nanoparticles, referred to in the main text and below as porous carbon (purchased from *FuelCellsEtc*).

Each device was clamped with O-rings to separate the input and output chambers (Fig. 2A of the main text). The proton-deuteron electrolyte was obtained by mixing a 0.15M HCl in H$_2$O solution with a D-electrolyte in different proportions. The latter consisted of 0.15M DCl (99% D atom purity) in $D_2O$ (99.9% D atom purity). Isotope fractions in the resulting electrolyte were prepared with an accuracy of ± 0.5%, and the volume of the electrolyte solution in the input chamber was of about a few $cm^3$. A Pt wire was placed inside a chosen $[H^+]:[D^+]$ electrolyte and served as the anode. The graphene membrane was the cathode (contacted with a microfabricated Au wire; see *fig. S1D*) and a dc voltage *V* was applied between it and the Pt wire. *Keithley's* SourceMeter was used both to apply the bias *V* and measure the current *I*. The gas flow and electric current were measured simultaneously. We checked each of our devices by a He leak detector to ensure that they were vacuum-tight. If no bias was applied, no flow of any gas through the graphene membranes could be detected. For more details, see ref. *5*.

For HD and $D_2$ having molecular masses 3 and 4, respectively, background fluctuations in our spectrometer were small enough to allow detection of the gas flow of less than $10^{-9}$ bar $cm^3$ $s^{-1}$, which translates into $\sim 10^{10}$ molecules per sec. For $H_2$ (mass 2) the resolution was much lower ($\sim 10^{-8}$ bar $cm^3$ $s^{-1}$ or below) because of remnant air and hydrocarbons present in the vacuum chamber. The lower resolution for $H_2$ made it necessary to use the largest possible graphene membranes and relatively large biases. Typically, we applied voltages around 1-2 V to produce hydron flows large enough to be detected with our mass spectrometer (see *fig. S4A*). Much smaller voltages were necessary to achieve detectable flows if porous carbon electrodes were used.

Note that at voltages above 1V water electrolysis begins to take place, which results in evolution of oxygen gas at the anode and hydron reduction at the graphene membrane. However, this has no consequences for the isotope separation results. Because the graphene membranes were impermeable to all atomic species, the gases detected by the mass spectrometer could only come from hydrons that cross graphene and recombine into hydrogen/deuterium at the output side. Therefore, the main effect of electrolysis in our case is to drive more hydrons towards the graphene membrane, which does not change the difference in $H^+$ and $D^+$ permeation rates. Furthermore, we note that, given that the absolute currents were of the order of µA (see, e.g., Fig 2A), the extra hydron concentrations due to the electrolysis can be estimated as $\sim 10^{-5}$ M, orders of magnitude below the 0.15 M acid concentrations used in our electrolytes. Also, the linear dependence of gas flow on *I*, which extended for both small (< 1V) and large (up to 20V) biases (*5*), confirms that electrolysis was only a complementary effect and does not affect the isotope separation. Finally, we note that the same separation factor as found by the mass spectrometry follows from our electrical measurements, where no electrolysis at the 2D membranes could possibly take place (especially for insulating hBN).

Determination of gas mole and atomic fractions in the output

Our mass spectrometer measured the gas flow, $\Phi_G$, in units of bar $cm^3$ $s^{-1}$, that is, $p_GV/t$ where $p_G$ is the partial pressure of a selected gas, *V* the detector's volume and *t* the time. Here we use subscript *G* to refer to different gases: $H_2$, HD and $D_2$. The number of molecules $N_G$ entering the mass spectrometer can then be calculated using the ideal gas law $p_GV = N_Gk_BT$ where $k_B$ is the Boltzmann constant and *T* the temperature (room temperature in our case). Therefore, $\Phi_G = N_Gk_BT/t$. On the other hand, the electrical current $I_G$ associated with the flow of charged nuclei is given by $I_G = 2eN_G/t$. The factor 2 takes into account that two hydrons, each carrying the elementary charge *e*, are required to create one gas molecule detected at the output.

If there is only one type of hydrons and all of them are successfully transferred through the graphene membrane, the mass and charge flows are related through the equation

$$\Phi_G = (k_B T/2e)\, I_G \tag{S1}$$

This equation is analogous to Faraday's law of electrolysis but the number of particles reacting at the electrode is given in equation (1) by the ideal gas law. More generally, the flow of each gas at the output side can be described by

$$\Phi_G = \gamma_G\, (k_B T/2e)\, I \tag{S2}$$

where $I$ is the total electric current and $\gamma_G \in [0,1]$ measures the proportion of the electrical current attributable to each gas at the output. For a given $[H^+]:[D^+]$ input, the above relations can be used (Fig. 2C of the main text) to determine the mole fraction of each constituent gas in the output flow as

$$[G] = \gamma_G / \{\gamma_{H2} + \gamma_{HD} + \gamma_{D2}\} \tag{S3}$$

If the mole fractions $[G]$ are known, the atomic fractions $[H]$ and $[D]$ in the output can be readily calculated. Because each $D_2$ molecule consists of two D atoms, HD of one H atom and one D atom and $H_2$ of two H atoms, the fraction of protium atoms in the output gas flow is

$$[H] = \{(\tfrac{1}{2})[HD]+[H_2]\}/\{[D_2]+[HD]+[H_2]\}$$

Similarly, the deuterium fraction at the output is given by

$$[D] = \{(\tfrac{1}{2})[HD]+[D_2]\}/\{[D_2]+[HD]+[H_2]\}$$

The above equations satisfy the obvious condition $[H] + [D] = 100\%$ for the output gas.

Reproducibility and accuracy of mass transport measurements

*Figs. S4B-D* show typical results of measuring the output flow for all three possible gases ($D_2$, HD, $H_2$) using different $[H^+]:[D^+]$ electrolytes at the input. One can see that the coefficients $\gamma_G$ depend strongly on $[H^+]:[D^+]$ concentrations. Furthermore, the plots illustrate the high reproducibility of our mass spectrometry measurements whereas the data scatter allows one to assess systematic errors. For example, *fig. S4B* shows measurements using the same device but different $[H^+]:[D^+]$ electrolytes in two different experimental runs, and *fig. S4C* is for the same $[H^+]:[D^+]$ inputs but two different devices. Measurements such as shown in *fig. S4* allowed us to calculate gas mole fractions in the output flow using equation (S3) for different $[H^+]:[D^+]$ inputs, and the results are presented in Fig. 2C of the main text.

Hydron permeation analysis

To relate the input $[H^+]$ and $[D^+]$ fractions to the output atomic fractions of different hydrons, we write

$$[H] = \{[H^+]P_H\}/\{[H^+]P_H + [D^+]P_D\} \tag{S4}$$

where $P_H$ and $P_D$ are the probabilities for protons and deuterons to cross the membrane, respectively. These probabilities are the same for electrical and mass-spectroscopy measurements. Both protons and deuterons can be expected to exhibit thermally activated (Arrhenius) permeation through 2D crystals – as reported previously for protons (*5*) – but with different activation energies $E_H$ and $E_D$, respectively. Therefore, we write

$$P_{H,D} \propto \exp(-E_{H,D}/k_B T) \tag{S5}$$

Therefore, equation (S4) takes the form

$$[H] = [H^+] / \{[H^+] + \exp(-\Delta E/k_B T)[D^+]\} \tag{S6}$$

This equation is used to analyze the measured input-output fractions in Fig. 2D of the main text. The fact that $\Delta E \approx 60$ meV (refs. *11,26*) is smaller than the activation energy of $\approx 250$ meV found in ref. *5* for Pt-activated graphene justifies the use of the Arrhenius dependence for deuterons, too.

In deriving equation (S6), we ignored for simplicity the fact that the attempt rates for protons and deuterons crossing the barrier could be different. Indeed, the frequency of zero point oscillations for protons is expected to be $\sqrt{2}$ times higher than for deuterons because the latter are twice heavier. If so, equations (S5-S6) should be modified with a pre-factor $1/\sqrt{2}$ in front of the exponent. However, in our analysis given in the main text, we have chosen to neglect possible differences in the attempt rates because it is known that they depend not only on zero point frequencies but often involve the oscillation of the neighboring atoms and, therefore, the attempt rates for hydrons often differ less than the simple model suggests (*26*). Moreover, even if we assume that the attempt rates differ by as much as $\sqrt{2}$, the consequences for our conclusions would be minor because of the exponential factor of ~10 is much larger than the pre-exponential one describing attempt rates. Indeed, we have found that if a pre-factor of $\sqrt{2}$ is introduced in equation (S6), the theory curve in Fig. 2B of the main text would bend slightly stronger but still remain within the error bars of the experiment. The good agreement between the experiment and theory in Fig. 2D probably indicates similar attempt rates for both hydron species (that is, the difference is less than $\sqrt{2}$).

Faradaic efficiency of hydron transfer

In our mass spectrometry measurements, if a pure proton electrolyte ([100% H$^+$]) was used at the input, we found $\gamma_{H2} \approx 1$ (inset of *fig. S5A*), in agreement with equation (S1) and the previous report (*5*). Because graphene is impermeable to atomic species, $\gamma_{H2} = 1$ means that all current-driven protons gain electrons at the output side of the graphene membranes where they evolve first into atomic and then molecular hydrogen (H$_2$). The same relation was observed for our reference devices that used porous carbon cloth instead of graphene (inset of *fig. S5A*). On the other hand, for 100% deuterons at the input ([100% D$^+$]) we measured $\gamma_{D2}$ of only $\approx 0.1$. Yet, for porous carbon we measured $\gamma_{D2} = 1$ (*fig. S5A*).

A reduction in Faradaic efficiency upon deuterium substitution was previously observed in electrochemical pumps based on, for example, high temperature proton conductive oxides (*25*). In that case, it was attributed to different mobilities of deuterons and protons in the oxides. In our case, H- and D- Nafion show the same bulk conductivity (see above), and the observed values of $\gamma_{D2}$ clearly show that some deuterons are converted into deuterium atoms at the input side of the graphene membranes. This process was sometimes evidenced through the formation of gas bubbles that became trapped in between the Nafion film and the gas-impermeable graphene membrane. Such bubbles were observed for the deuteron input if large current densities were applied for long periods of time to small-area devices made from exfoliated graphene (see *figs. S6B,C*). We believe that, in this case, the slower permeation rate for deuterons through graphene led to their accumulation at the inner face of graphene membranes where they eventually evolved into D$_2$. Although the rate of deuterium/protium evolution reaction on graphene is expected to be low (*36,37*), the slow permeation of deuterons through graphene probably results in in the situation that the evolution and permeation rates become comparable. It is important to emphasize that no bubbles were observed for large-area CVD devices, where current densities were much smaller, or for our electrical measurements. This indicates that the bubble formation is a background process occurring only under high current densities and, below a certain threshold, can be compensated by reabsorption of deuterium back into the Nafion film.

For completeness, *fig. S6A* shows the relation between charge and mass flows for intermediate concentrations of protons at the input. By adding $\gamma_G$ for $D_2$, HD and $H_2$ using the same input electrolyte, we can determine the total gas flow $\Phi_G = \Sigma\gamma_G(k_BT/2e)I \equiv \Gamma\times(k_BT/2e)I$ where $I$ is the total electric current and $\Gamma = \gamma_{H2} + \gamma_{HD} + \gamma_{D2} \in [0,1]$. If $\Gamma = 1$, all the electric current translates into the mass flow (100% Faradaic efficiency). If $\Gamma <1$, some of the hydrons reaching the graphene membrane accept electrons from it but do not pass through. Note that $\Gamma$ characterizes the transparency of membranes with respect to different hydron species and, naturally, is expected to be lower for deuterons that experience a higher activation barrier. However, $\Gamma$ is not directly connected with the isotope separation factor $\alpha$.

It is worth mentioning that in principle a difference in production of protium over deuterium can occur even in the absence of a barrier film. This may be, for example, due to different gas evolution rates for different hydrons at the electrodes (*19,38-42*). No such effects were detected in our control experiments with porous carbon electrodes (*figs. S5B-C*), which shows that the presence of 2D crystals is essential for the reported isotope effect. The absence of any detectable separation in the control experiments with porous carbon electrodes is hardly surprising. First, the isotope effect reported for electrolysis using polymer electrolytes such as Nafion is small, exhibiting a separation factor $\alpha \approx 3$ even under optimum conditions (*38*). Such $\alpha$ would lead only to differences in output [H] fractions within the error bars of *fig. S5C*. Second, electrolysis is sensitive to applied voltages and current densities and requires fine tuning and cleaning of electrodes in order to achieve the above modest separation factor (*38-42*). In our experiments, no special preparation of electrodes was necessary, and currents/voltages could be significantly different for different devices. This comparison also emphasizes the fact that, in addition to the large separation factor of 10, graphene membranes provide a very robust isotope effect.

Separation factor

The efficiency of isotope separation techniques is characterized by a separation factor (*15*)

$$\alpha = \{[H]/[D]\}/\{[H^+]/[D^+]\}$$

which is the ratio of relative concentrations of protons and deuterons at the input and output sides of a separation device. If the concentration of one of the isotopes is low, it is straightforward to show that equation (S6) leads to

$$\alpha = \exp(\Delta E/k_BT) \tag{S7}$$

Using the isotopic shift energy $\Delta E \approx 60$ meV, equation (S7) yields $\alpha \approx 10$ at room *T*. Note that exactly the same exponent defines the ratio of hydron conductivities, $\sigma_H/\sigma_D$, in electrical conductivity measurements as follows from equation (S5). The experimentally found ratio $\sigma_H/\sigma_D = 10\pm0.8$ yields the same $\alpha$ as the mass-spectrometry measurements and agrees well with the above value of $\Delta E$ found from optical spectroscopy.

Mass transport using CVD graphene

To fabricate cm-size devices for mass-spectrometry measurements, we used CVD graphene grown on copper (provided by *BGT Materials*). One side of the copper foil was coated with a thin layer of PMMA, and the other side was etched in oxygen plasma to remove graphene from this side of the foil. The copper was then etched using the standard ammonium persulfate solution. The remaining graphene-PMMA film was thoroughly cleaned in deionized water and transferred onto a Nafion 1110 film purchased from *FuelCellsEtc*. The assembly was baked in a humid atmosphere at 130°C (*5*) and glued with epoxy over a cm-size hole made in a rubber sheet (*figs. S1E-F*) which also served as a gasket to separate the liquid cell and the vacuum chamber (Fig. 2A of the main text). Next, the PMMA was dissolved in acetone/hexane, avoiding contact of the solvents with the opposite side of the Nafion film. In the final assembly, CVD graphene is clearly visible on top of Nafion. Its inspection in a high-magnification optical microscope revealed some folds and cracks that occupied ≈1% of the membrane area. Typical resistivities of the transferred CVD graphene were ≈1 kOhm per square, again indicating high quality of the transfer from copper onto Nafion. Finally, CVD graphene was decorated with Pt nanoparticles as described above and electrically contacted using silver epoxy. The CVD devices were measured using mass spectrometry in the same way as those made from exfoliated graphene crystals.

*Figure S7* shows that our CVD graphene membranes exhibited essentially the same behavior and the same high selectivity as the micron-size mechanically exfoliated graphene crystals (cf. Fig. 2 of the main text). For an electrolyte input containing only deuterons [100% $D^+$], we again found $\gamma_{D2} \approx 0.1$ whereas for a [100% $H^+$] input, $\gamma_{H2}$ was ≈1 (*fig. S7A*). For an input of [50% $H^+$: 50% $D^+$], that output flow contained ≈90% of H (*fig. S7B*), which is only slightly different from the composition observed for our micron devices and within the experimental error bars in Fig. 2D of the main text.

It may come as a surprise that the isotope separation is possible if cracks are present in graphene membranes. Indeed, cracks in CVD graphene allow electric contact between Nafion on the opposite sides of the graphene membrane. Therefore, in conductivity measurements such as in Fig. 1 of the main text, the largest contribution to the measured current will come from protons moving through Nafion in the cracks, rather than through the significantly less proton-conductive graphene. Accordingly, CVD graphene could not be used in our hydron conductivity measurements. The situation is radically different in the mass spectrometry setup that effectively presents an electrochemical pump (*25,26*). In this case, hydrons are driven (pumped) only through the graphene areas that are electrically contacted and, therefore, cracks contribute relatively little. The slightly lower separation factor observed for CVD graphene could be due to those cracks, although the difference is minor and within the experimental error.

Potential applications

Hydrogen isotopes are important for various technologies, particularly, in analytical and tracing analysis (*43-46*) and nuclear energy applications (*15,16,47*). Accordingly, it is not surprising that a large number of methods have been developed to separate hydrogen isotopes (*11-21,47,48*). These methods include liquid $H_2O$ distillation ($\alpha \approx$ 1.05), electrolysis ($\alpha \approx$ 2 to 10), ammonia-hydrogen exchange ($\alpha \approx$ 3 to 6), liquid $H_2$ distillation ($\alpha \approx$ 1.5), water-hydrogen exchange ($\alpha \approx$ 2.8 to 6), aminomethane hydrogen exchange ($\alpha \approx$ 3.5 to 7), water hydrogen sulfide exchange ($\alpha \approx$ 1.8 to 3), quantum kinetic isotope separation ($\alpha \approx$ 6 at liquid nitrogen $T$) and multi-photon laser separation ($\alpha$ >20,000). All these methods, except for the multi-photon laser separation that is deemed impracticable (*16*), are or were used in industrial-scale heavy water production. Our report shows that graphene and hBN membranes offer a highly competitive separation factor $\alpha \approx 10$ with potentially a simple setup and no other chemical compounds but water at the input. Similar considerations apply for separation of tritium and its removal from heavy water (*16*). Using the known proton-triton energy shift $\Delta E \approx 88$ meV for hydrogen bonds in Nafion/water (*49*), equation (S7) yields $\alpha \approx 30$ and $\approx 3$ for protium-tritium and deuterium-tritium separation, respectively.

Furthermore, it is instructive to estimate possible energy costs associated with the described isotope separation method. For Pt-activated graphene, its proton conductivity $\sigma$ is $\approx 0.1$ S cm$^{-2}$ (*5*). Using low voltages $V \approx 0.1$ V, we can achieve proton currents $I = \sigma V \approx 100$ A per m$^2$. This translates into the H$_2$ production rate $R = I/2N_A e \approx 2$ moles per hour per square meter (where $N_A$ is the Avogadro number) and yields the energy costs $IV/R = 2N_A eV \approx 5$ Wh per mole or $\approx 0.3$ kWh per kg of feed water. Also, note that, according to the Fenske equation, the large $\alpha$ implies only a few stages of enrichment for cascade plants to obtain 99% pure heavy water. In principle, throughputs ~0.1 kg m$^{-2}$ per hour can be achieved using Pt-activated hBN that exhibits 30 times higher proton conductivity ($\sigma \approx 3$ S/cm$^2$ as discussed above) but CVD growth of hBN monolayers in industrial quantities has not been demonstrated yet.

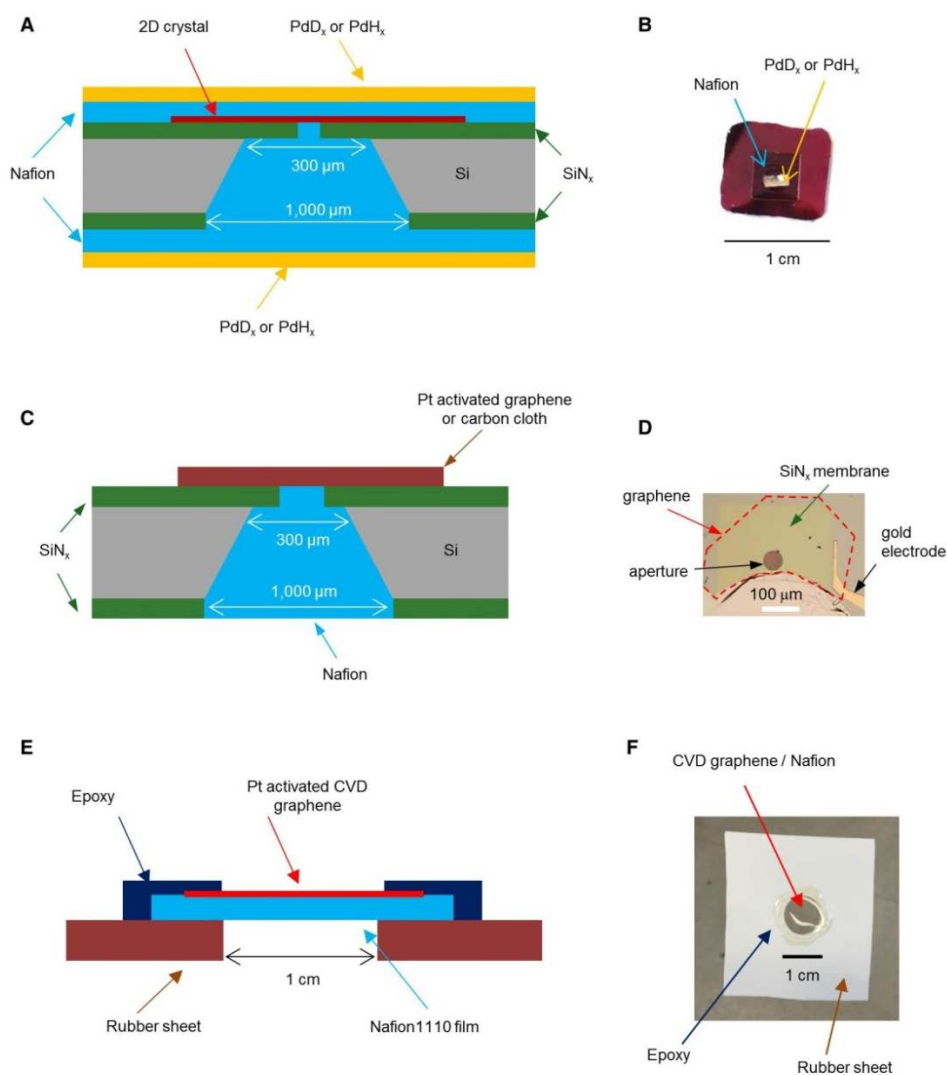

**Fig. S1.**
Device fabrication. (**A**) Devices for measurements of electrical conductivity. A 2D crystal is suspended over a hole etched into a free standing silicon-nitride (SiN$_x$) membrane. Both sides are coated with Nafion, and Pd electrodes are attached mechanically. (**B**) Optical photo of the final device for electrical measurements. (**C**) Schematic of mass spectrometry devices. For control experiments, a carbon cloth was used instead of graphene. (**D**) Optical image of one of our devices (view from the output side). A graphene monolayer (its position is outlined by the dashed lines) covers a circular aperture that is etched in a silicon-nitride membrane visible as a yellowish square. Graphene is electrically contacted using a gold electrode. Scale bar, 100 µm. The bottom area seen as beige is an adjacent multilayer graphene flake. (**E**) Schematic of CVD-graphene devices used for mass spectrometry measurements. CVD graphene is transferred onto a Nafion film that is in turn glued to a gasket using epoxy. (**F**) Optical photo of an assembled device.

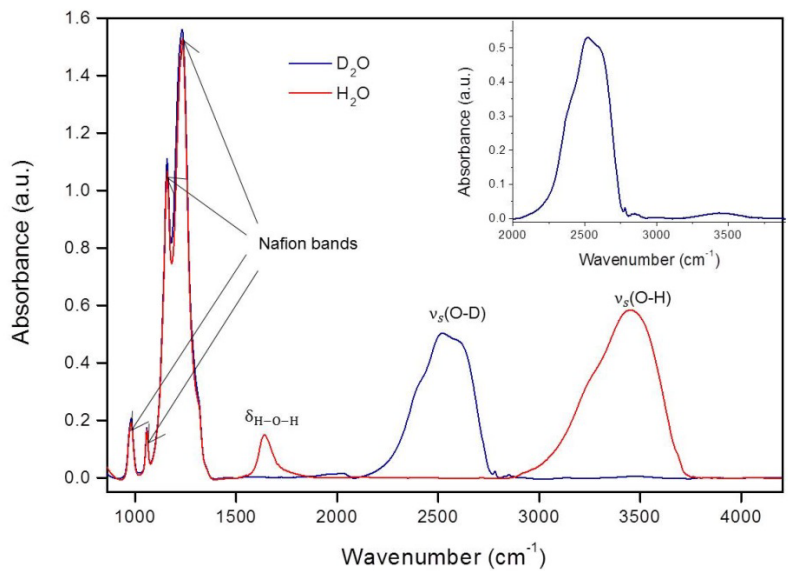

**Fig. S2**

Infrared spectroscopy of H- and D- Nafion. IR spectra of Nafion films exposed to light and heavy water. Inset: Magnified D-Nafion spectrum shows a small OH peak at 3,500 cm$^{-1}$ due to remnant protium.

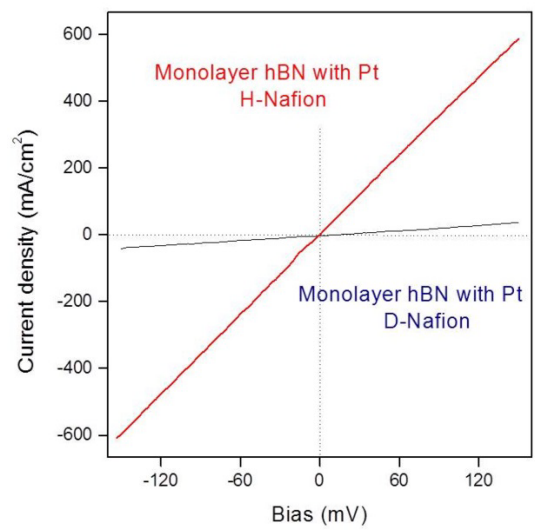

**Fig. S3**
I-V responses for Pt-decorated monolayer hBN.

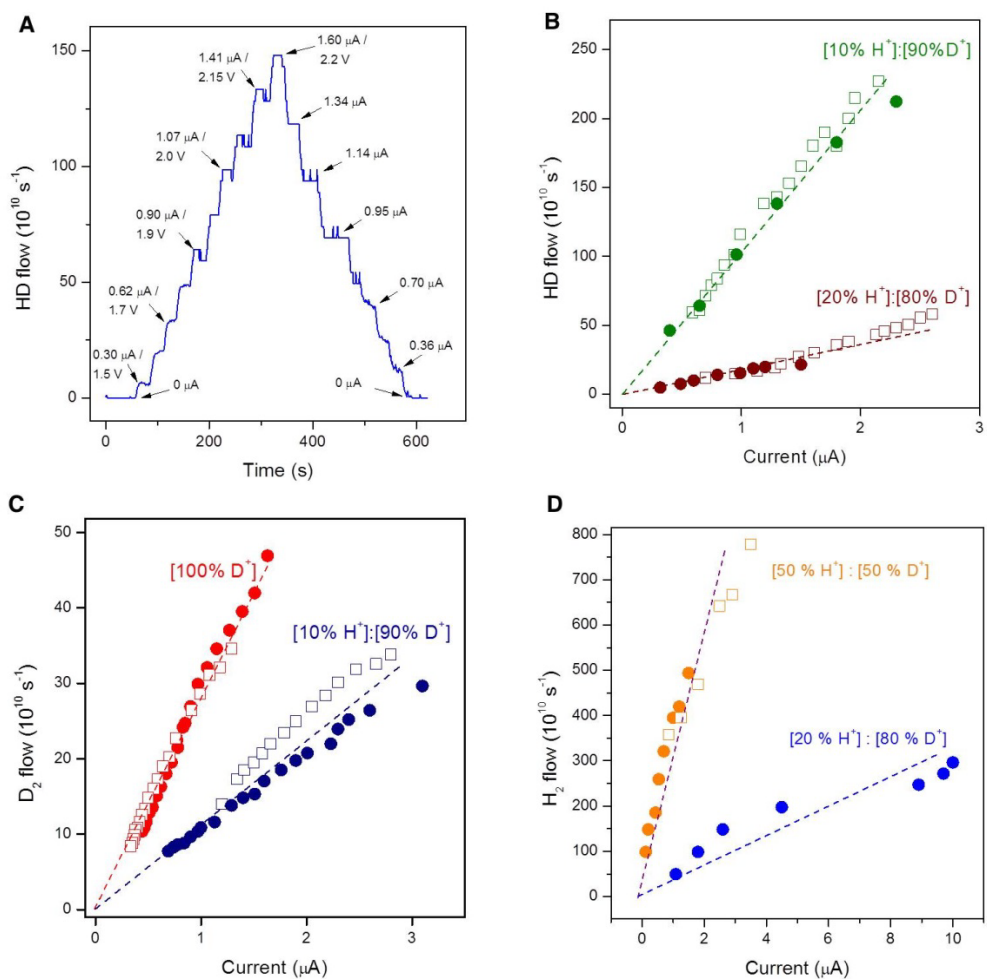

**Fig. S4**

Mass transport measurements using exfoliated graphene. (**A**) Typical data-acquisition run used in measurements of hydron transport. The particular example is for an HD flow (atomic mass 3) through graphene using a [10% $H^+$]:[90% $D^+$] mixture at the input. (**B**) Four different runs for the same device (each data-acquisition run looked similar to that shown in *fig. S4A*). Squares and circles represent different runs for the same [$H^+$]:[$D^+$]. (**C**) Two different devices (squares and circles) using the same [$H^+$]:[$D^+$] inputs. The measured gas is $D_2$. (**D**) Three different runs using another device for different [$H^+$]:[$D^+$] inputs. The measured gas is $H_2$.

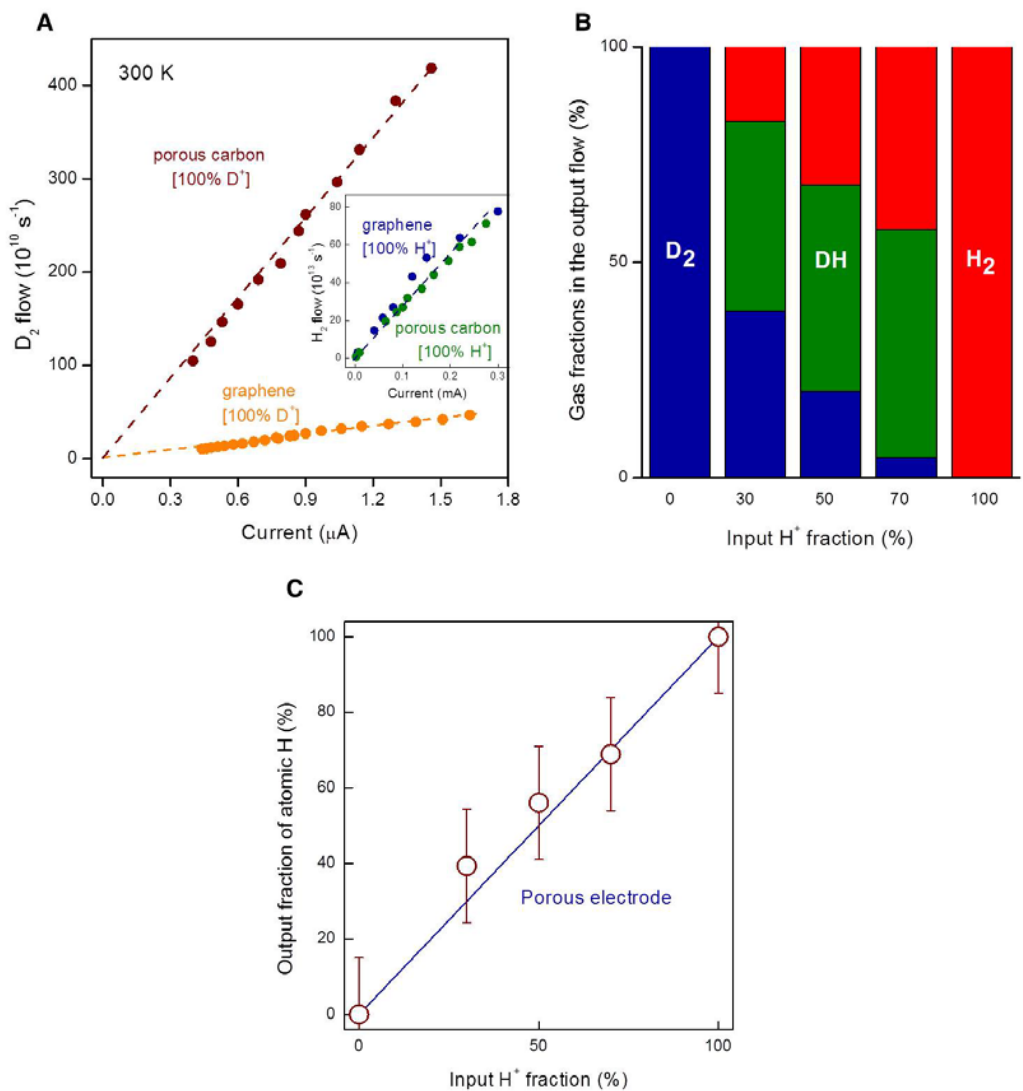

**Fig. S5**

Control mass-transport experiments using a porous carbon electrode. (**A**) $D_2$ flow detected using Pt-activated graphene membranes (orange symbols) and porous carbon (brown) with only deuterons at the input. Inset: Same for a 100% proton input. (**B**) Gas fractions for different $[H^+]:[D^+]$ inputs using porous carbon. (**C**) Atomic output versus input calculated using the data in (**B**). The blue line shows the behavior expected for the case of no selectivity: $[H] \approx [H^+]$.

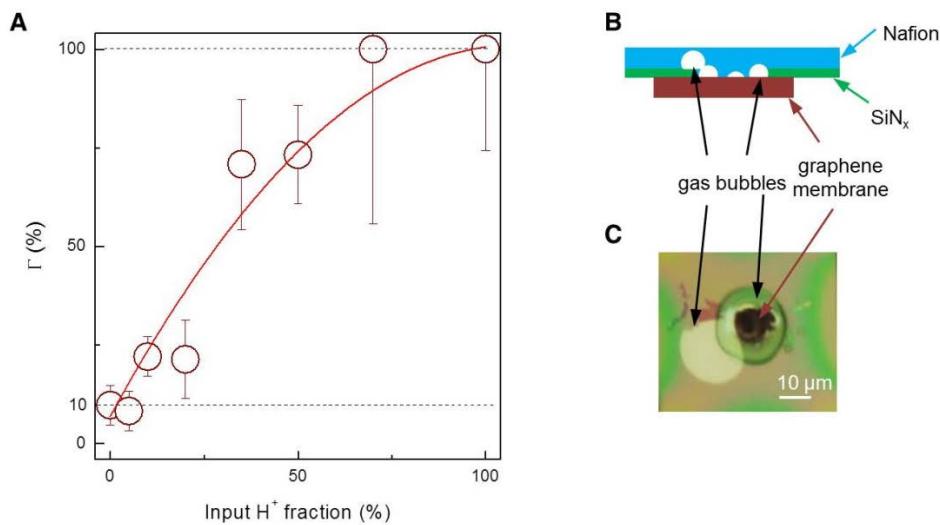

**Fig. S6**

The fraction Γ of hydrons converted into the gas flow at the output side of the graphene membrane cathode. (**A**) Γ for different proton concentrations [$H^+$] in the input electrolyte. The red curve is a guide to the eye. (**B**) Schematics of the bubble formation. (**C**) Optical image of a device that during its final run was measured at high currents and using a [100% $D^+$] electrolyte. The top view is from the input side covered with a Nafion film. It is optically transparent and the green fringes appear due to Nafion areas of different thicknesses.

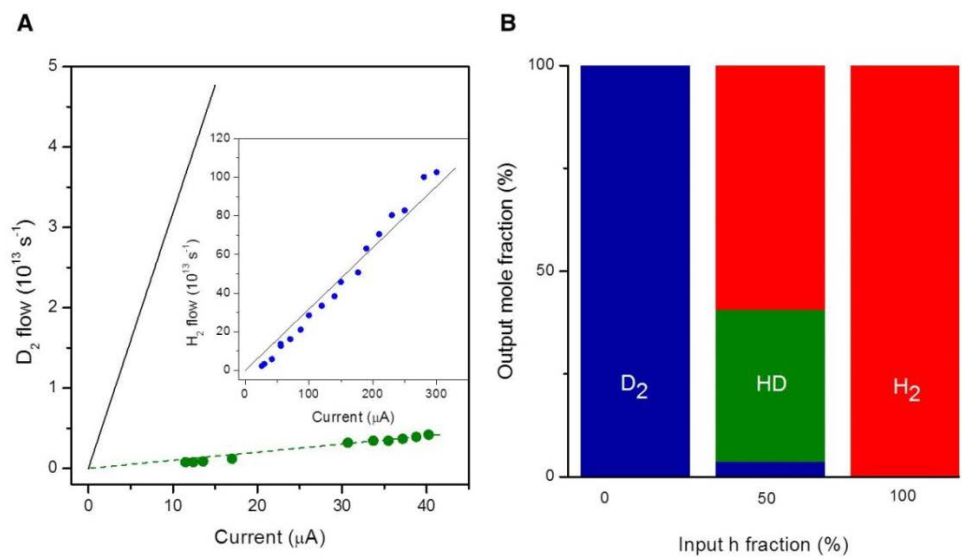

**Fig. S7**

Isotope separation using CVD-graphene membranes for electrochemical pumping. (**A**) Flow-current characteristics for only deuterons (main panel) and only proton (inset) in the input electrolyte. Main panel: for 100% deuterons, we again observe $\gamma_{D2} \approx 0.1$ as for the case of exfoliated graphene; the black line shows $\gamma = 1$. Inset: for a [100% H$^+$] input, $\gamma_{H2} \approx 1$ (black line shows $\gamma = 1$). (**B**) Output gas composition for three different input concentrations using CVD-graphene devices.